# Schema Evolution in Interactive Programming Systems

Jonathan Edwards[a] , Tomas Petricek[b] , Tijs van der Storm[c,d] , and Geoffrey Litt[e]

a   Independent, Boston, MA, USA

b   Charles University, Prague, Czechia

c   Centrum Wiskunde & Informatica (CWI), Amsterdam, Netherlands

d   University of Groningen, Groningen, Netherlands

e   Ink & Switch, USA

**Abstract**    Many improvements to programming have come from shortening feedback loops, for example with Integrated Development Environments, Unit Testing, Live Programming, and Distributed Version Control. A barrier to feedback that deserves greater attention is *Schema Evolution*. When requirements on the shape of data change then existing data must be migrated into the new shape, and existing code must be modified to suit. Currently these adaptations are often performed manually, or with ad hoc scripts. Manual schema evolution not only delays feedback but since it occurs outside the purview of version control tools it also interrupts collaboration.

Schema evolution has long been studied in databases. We observe that the problem also occurs in non-database contexts that have been less studied. We present a suite of challenge problems exemplifying this range of contexts, including traditional database programming as well as live front-end programming, model-driven development, and collaboration in computational documents. We systematize these various contexts by defining a set of layers and dimensions of schema evolution.

We offer these challenge problems to ground future research on the general problem of schema evolution in interactive programming systems and to serve as a basis for evaluating the results of that research. We hope that better support for schema evolution will make programming more live and collaboration more fluid.

## The Art, Science, and Engineering of Programming

## 1 Introduction

Many improvements to the practice of programming have come from speeding up feedback loops. Replacing punched cards with interactive terminals was one of the greatest leaps of productivity in the history of programming. Unit tests and interactive type checking provide faster feedback on errors that modern programmers now expect. Research on Live Programming [55] seeks to provide immediate feedback on runtime behavior while editing. Feedback is equally valuable in collaborative programming [12, 28, 42, 56] to convey changes between coworkers with less friction. However there is still much room to improve feedback loops throughout software development.

In this paper we identify a common problem, *Schema Change*, that interferes with feedback loops in diverse contexts of software development. Schema Change has been extensively studied in the context of databases where it necessitates manual and error-prone data migrations [53]. We observe analogous problems occuring in varied and less-studied contexts. For example, many live programming techniques treat state as ephemeral and recreate it after every edit, but when the shape of longer-lived state changes then the illusion of liveness is shattered – hot reloading works until it doesn't. Likewise collaborative programming works smoothly when code changes can be exchanged that do not alter the expected shape of external data, but changes that cross that line must be coordinated outside of the version control system. We discuss further examples in Section 9. In general, when the shape of data becomes a dependency, changes to that data's shape can require manual intervention which interrupts feedback loops. To coordinate research on these related problems we shift the focus to *Interactive Programming Systems* [37] and generalize the term 'schema' to include not just database schema but also data types and specifications of all sorts, whether expressed in a formal language or left implicit in code.

Our primary contribution is a suite of challenge problems exemplifying a diverse range of typical schema change scenarios that interrupt feedback loops. We believe these problems have not been completely solved in theory or practice. While some of them come from our individual research efforts we have attempted to make neutral problem statements independent of our preferred approaches. None of our ongoing efforts yet address the full range of problems. Indeed, the aim of this paper is to provide a basis for collaboration between such related efforts, facilitate further research, as well as to serve as a basis for evaluating research results. We invite contributions of new challenge problems and look forward to discussing competing solutions.

## 2 Layers and Dimensions of Schema Evolution

To paraphrase Stewart Brand [10], "almost no programs evolve well. But all programs (except for monuments) evolve anyway, however poorly, because the usages in and around them are changing constantly." Programs evolve both while they are being created and during their operation in response to new needs. As with buildings, different aspects of programs evolve at different paces (Figure 1). And "because of the different rates of change of its components, a program is always tearing itself apart."

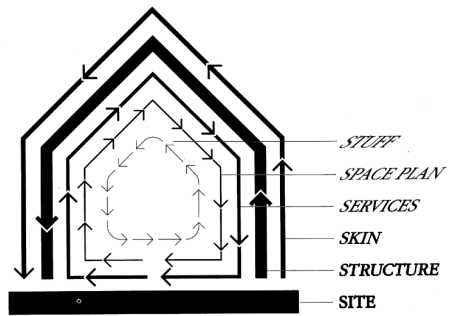

■ **Figure 1** Pace layers from Stewart Brand's *How Buildings Learn* [10]. Different layers of a building change at different pace. The geographic site of a building is eternal, its structure remains stable for decades, space plan changes every couple of years.

In both buildings and programs, making a change at surface layers (skin or stuff) is relatively easy, but a change at a more fundamental layer (structure) has cascading effects on layers that depend on it. In the case of programs, we refer to the more fundamental layer as *schema*, although we do not use the term in a precise technical sense. Schema are what define the shape of data and code, either as database schema, data types declared explicitly in code and checked statically, dynamically checked types and specifications, and even expectations implicit in the code. The key property of schema for us is that they constrain the shape of other layers of a program, namely data and code. As a result, a change of schema requires a corresponding change in data and code.

The situation is illustrated on the right. We start with code $C_1$ and data $D_1$ that have a shape corresponding to schema $S_1$. When the schema changes from $S_1$ to $S_2$ ($\rightarrow$), a corresponding change needs to take place in data $D_1$ and code $C_1$. The problem of *schema evolution* is synthesising ($\Rightarrow$) suitable corresponding transforma-tions that will update data $D_1$ and code $C_1$ into new versions $D_2$ and $C_2$ that correspond to the new schema $S_2$. The complexity of the problem varies and finding a suitable transformation may require user input. If there is code, it typically needs to be transformed, but data can sometimes be discarded. We refer to such transformations done by a single user as *local* and draw them inside a box.

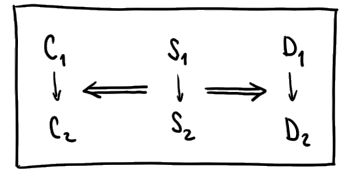

Collaboration adds the *non-local* dimension to the problem. A program may have multiple variants that exist independently and changes made to them may need to be synchronized. The situation is illustrated on the right. Here, the initial schema $S_1$ is transformed by two users independently into schema $S_2$ and $S_3$. The corresponding data is transformed accordingly, produc-ing $D_2$ and $D_3$. The challenge is now merging these two diverging transformation and obtaining a schema $S_4$ and data $D_4$ that adopt both of the changes.

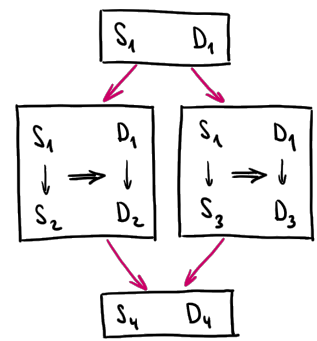

The difficulty of the problem depends on what kind of synchronization we want to support. If all variants of the program should eventually adopt all changes (converge), the complexity is lesser. If we allow users to apply only certain changes from other users (divergence), the complexity is greater. A common case may be a combination of the two where types and schema of a program converge, but data may diverge.

**Locality of Change**   We can view diverse real-world schema evolution problems from a common perspective by distinguishing between two different dimensions of schema change. The *local dimension* is concerned with how schema evolution affects data and code within a single program. The *non-local dimension* is concerned with how schema evolution propagates across different program variants that may exist independently.

**Program Layers**   We consider the more permanent layer of *schema* and two less permanent layers of code and data that are structured by those schema. In specific case studies, the layers may map to different aspects of programs, some may be not present, and some may be implicit.

- *Schema* – structures program data and determines aspects of code. A change in schema requires a corresponding change in data and code. Schema may be explicit (database schema, type declarations) or implicit (expected shape of data structures or a document).
- *Data* – information stored by the program. Data may be more permanent (data in a database) or less permanent (state in live programming). When schemas evolve, data needs to be updated or (in case of transient state) discarded.
- *Code* – or program logic, but not including source code that defines schema. When schemas evolve, code needs to change to work with data of the correct shape.

**Program Variants**   In programming systems that enable collaboration, multiple variants of a program may exist concurrently and their schema, data and code may need to be synchronized. Although collaboration is missing from some of our case studies, it adds an important dimension that is crucial for schema evolution in future programming systems. We consider two characteristics.

- *Convergence* vs. *Divergence*. In the convergence model, all program variants eventually adopt all changes. It may not be possible to adopt a change without adopting all earlier changes. In the divergence model, a user may choose only particular changes they want to adopt, but keep other aspects of their custom design.
- *Centralized* vs. *Decentralized*. In the centralized model, changes (schema evolution) can only originate from a particular source. In the decentralized model, changes can be done on any of the multiple co-existing variants of a program.

**Challenge Problems**   These dimensions, layers, and variants generate a wide range of problem configuration. Our aim is not to cover all of them. Instead we use this perspective to recast a number of existing challenges in real-world programming systems as instances of the same problem of schema evolution. The case studies show how well-known solutions or problems in one domain map to another domain. We hope they provide an inspiration and a benchmark for developers of future systems.

### 3  Live Programming for the Elm Architecture

The first challenge problem that we present is live programming of user interfaces based on the Elm architecture. In this model, a reactive web application is structured in terms of current state and events that affect the state. The implementation then consists of two functions called update and render:

```
1  type State = { .. }
2  type Event = .. | ..
3
4  val update : State -> Event -> State
5  val render : State -> Html
```

The schema involved here are statically checked Elm types. The programming model works as follows:

- State represents the application state, i.e., everything that the user can work with.
- Event represents events that the user can trigger by interacting with the application.
- update is called when an event happens and computes a new program state.
- render takes the current state and produces a visual representation of the page.

As an example, consider the well-known todo list app. Its state consists of a list of items, each with a unique ID, a title and a flag indicating whether it has been completed. The events represent changing of an item, deletion and addition:

```
1  type Item = { id : id; title : string, completed : bool }
2  type State = { items : Item list }
3  type Event =
4      | SetTitle of id * string
5      | SetCompleted of id * bool
6      | Remove of id
7      | Add of string
```

Some systems that use the Elm architecture support hot reloading when the implementation of the render or update function changes, but they typically discard state and restart the application when the State type changes. A more effective live programming system would allow live updates to the structure of the two types.

**Challenge #1: Live State Type Evolution**

Assume that we have a running todo list with the above state and events. To test the application, the programmer has already created a number of items and so there is a value of the State type that represents the current state of the application such as:

```
1  { items : [
2      { id = 1; title = "Check Twitter"; completed = true }
3      { id = 2; title = "Write the paper"; completed = false } ] }
```

**Table 1** Changes to types and how a programming system should handle them

| Change | What | How |
|--------|------|-----|
| Add | case to Event | Requires adding corresponding case to update |
| Remove | case from Event | Remove unused code from update |
| Add | field to State | Migrate state value and initialize field |
| Remove | field from State | Migrate state (assuming field unused) |
| Modify | structure of State | Migrate state value and edit code accordingly |

There is a number of edits to the types that the programmer may now want to do without restarting the application and losing the data. For example, they may want to change the completed field of Item to instead store optional completion time:

```
1  type Item = { id : id; title : string; completed : Maybe<DateTime> }
2  type State = { items : Item list }
3  type Event = .. | SetCompleted of id * Maybe<DateTime> | ..
```

To apply the change without restarting the application, the existing data need to be migrated to match the new type definition. In this case, this likely cannot be done fully automatically and the programmer may need to provide mapping (false to Nothing, true to Just(DateTime(2024,5,1))). If the programmer changes Item and SetCompleted at the same time, the update function likely does not need to change, but the rendering code needs to be modified (the programmer needs to specify how to map DateTime option back to a Boolean for a checkbox).

More generally, there is a number of edits to the types that the programmer may want to do without restarting the application. The edits and the desired handling in the programming system are summarized in Table 1. Modifying the Event type does not require data migration (assuming that the change does not happen when there are unprocessed events). It may result in unused code or missing cases in update that need to be addressed. Modifying State ranges from relatively simple problems (e.g., adding a new field with a default value) to challenging case. In particular, if the programmer refactors State to a semantically equivalent type, the programming system should, in principle, be able to automatically migrate both the original value and implementation of both functions to match the new type.

**Primitive Schema Transformations**     Here, the evolving schema is the State type. When the type evolves, the program logic (code) and the current live value (data) needs to evolve correspondingly. The challenge is easier if we see the evolution as a sequence of more primitive transformations, such as those in Table 1, that each has a corresponding primitive code and data transformation. The primitive transformations may either be invoked through a UI, or inferred from textual code edits made by the programmer.

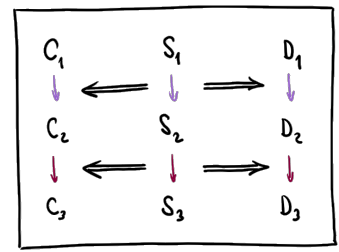

| oid | item | quantity | ship_date | customer_name | customer_address |
|-----|------|----------|-----------|---------------|------------------|
| 1 | Anvil | 1 | 2/3/23 | Wile E Coyote | 123 Desert Station |
| 2 | Dynamite | 2 | | Daffy Duck | White Rock Lake |
| 3 | Bird Seed | 1 | | Wile E Coyote | 123 Desert Station |

**(a)** Spreadsheet recording orders with all associated information

| cid | customer_name | customer_address |
|-----|---------------|------------------|
| 1 | Wile E Coyote | 123 Desert Station |
| 2 | Daffy Duck | White Rock Lake |

**(b)** Customers table after schema evolution

| oid | item | quantity | ship_date | cid |
|-----|------|----------|-----------|-----|
| 1 | Anvil | 1 | 2/3/23 | 1 |
| 2 | Dynamite | 2 | | 2 |
| 3 | Bird Seed | 1 | | 1 |

**(c)** Orders linking to Customers, after schema evolution

■ **Figure 2** Schema and data before and after performing Extract Entity

**Remarks: Requirements and Implementation**

The key requirement from the live programming perspective is to "do minimal harm" to the experience of liveness. The migration of *data* needs to be done automatically and the system should strive to produce a new state that is as near as possible to the previous state. Migrating *code* can be done in various ways, but it is desirable to avoid breaking the render function as this would make the new application state impossible to visualize.

The specific case where the type is refactored to a semantically equivalent type is related to the Extract Entity challenge discussed below. Finally, a programming system tackling this challenge may rely on structure editing [68] so that the system has access to a high-level logical description of the edits performed by the user.

## 4 Entity Evolution in Data-Centric Systems

If we see programming systems as stateful environments where programs co-exist with their persistent data, we can gain valuable insights from research on database systems. Although schema evolution is well-studied in databases (Section 9), such past work does not resolve all problems in our second challenge.

As an example, consider the Acme Corporation that records orders from various customers for various products. They use a spreadsheet (Figure 2a) with a row for each order and columns for information about the customer and product. The shipping department filters this table on blank ship dates to see what they need to ship. But after a while the orders department realizes they are wasting effort duplicating the address for new order from an old customer. And when the customer's address changes they have to go back and edit all of their orders.

■ **Table 2**  Changes to schema and how a programming system should handle them

| Change | What | How |
|--------|------|-----|
| Extract | new type of entity | Migrate data to a new table and add links |
| Absorb | a linked type of entity | Migrate data and remove unneeded table |
| Merge | multiple data items | Automatic de-duplication with mistake handling |
| Split | multiple data items | Interactively, following user guidance |

**Challenge #2: Extract Entity**

Acme needs to evolve the above schema and data into two tables (Figure 2b, 2c), one with orders that links to one with customers. We refer to this as the *Extract Entity* operation, because it introduces a new *entity*, a referencable holder of mutable attributes. The operation is needed when we realize that some attributes of one type of entity should belong to a distinct type of entity that will be associated with the first one.

The system needs to provide a mechanism for referencing entities, such as by using unique identifiers (here consecutive numbers). This allows, for example, a spelling error in a customer address to be fixed in one place. It also allow a customer name or address to change without breaking the connection from all associated orders. Unique identifiers are the essence of entities, whether they are uniquely generated primary keys in a database or an abstraction of a memory address in a programming language. As spreadsheets do not support relationships between entities, Acme will need to migrate their data to a database and restructure the original flat table.

The *Extract Entity* operation must 1) merge duplicates, 2) assign unique identifiers, and 3) reference these identifiers from the orders. Our example took the simple approach of merging entities whose attributes are all equal. But that might not be correct in all cases. What if there was a typo in one of the addresses that made it look different from other instances of the customer? To repair that error we need to merge the falsely distinct customers into one and fix any references from orders. We call this operation *Merge Entities* – note that it is not a schema change but a data change, yet nevertheless not commonly supported as an operaton in database APIs or DMLs.

Entity evolution can happen through operations such as those listed in Table 2. Every schema operation should have an inverse as the change in requirements that triggered an evolution might change back again. Invertability features in other work on schema evolution and *bidirectional transformations* [13, 26, 31]. The inverse of *Extract Entity* is *Absorb Entity*. It raises the question of how to handle *one-to-many* relationships. For example should absorbing orders into customers do a relational join, returning us back to the starting point, or should it nest orders within customers, as a NoSQL database might do?

The inverse of *Merge Entities* is *Split Entity*, but that can be awkward. Imagine that we didn't have addresses for customers to disambiguate them with and only knew their non-unique names. Then *Extract Entity* will conflate them and discard the information needed to properly invert the operation. Should this information be retained? One could object that we shouldn't have taken orders for people we can't uniquely identify yet. And even if that was necessary it means the customers really aren't distinct entities and shouldn't have been extracted. These are arguments that we shouldn't need *Split Entity*, contradicting the principle that evolution operations should be invertible. It is interesting to observe that evolution in nature can also fail to invert, leaving behind vestigial features. We consider this to be an open issue.

**Remarks: Representation and Replication**

The central problem is to coordinate the schema evolution with code edits to keep the system executing live and correctly. Solutions may infer the desired schema evolution by analysing source code edits, but would typically offer UI affordances for the commands. They should try to minimize the number and complexity of commands that must be introduced and also the interruption to live execution of the code.

We can also consider the Extract Entity challenge in a local-first [39] case, where the context is an application that offers peer-to-peer collaboration. In this context the schema change does not need to be performed interactively – a developer can take some time to build, test, and package a solution, perhaps using an API or DSL. The hard part comes when the schema change is to be deployed across all the replicas. This deployment must synchronize code and data upgrades (or decouple them as in Cambria [46]). The deployment must also deal with migrating "in flight" operations so that all replicas converge on the same state without data loss, and ideally without centralized coordination. A radical approach could try to incorporate schema change operations into the underlying datastore itself (possibly a CRDT) and integrate code as well, but layered solutions are welcome too.

**Challenge #3: Code Co-evolution for Extract Entity**

In a programming system, there is typically also some code for programmatic manipulation of entities. The code can, for example, implement a simple CRUD UI that also provides the Acme shipping department its view of unshipped orders. The holistic perspective of programming systems lets us see this code not as separate from the system, but as its integral part. The code thus also needs to co-evolve when the entity evolves. An example is a SQL query to report all pending orders in the original schema:

```
1  SELECT order_id, item, quantity, customer_name, customer_address
2  FROM Orders WHERE ship_date IS NULL;
```

After applying the *Extract Entity* operation as described earlier, the code should be rewritten into the following query that joins the two linked tables:

```
1  SELECT order_id, item, quantity, customer_name, customer_address
2  FROM Orders
3  JOIN Customers ON Orders.customer_id = Customers.customer_id
4  WHERE ship_date IS NULL;
```

In database systems, the system can use a mechanism such as a view and execute the original query over the view. When applying code migrations, the system should minimize the need for users to manually intervene, which interrupts feedback from the migration. It should also minimize the possibilities that developer-written code is subtly incorrect in edge cases, preferably by reducing the need for developer-written code in the first place.

> **Equivalence in Schema Evolution**    In database systems, the schema $S_1$ changes into $S_2$ and the data $D_1$ need to transform to corresponding $D_2$. Our framing shows that queries $Q_1$ also need to transform, which is not always addressed in work on schema evolution. Extract/Absorb Entity are also interesting because they transform an equivalent pair of schemas and so it should be possible to apply them (and the corresponding data and query transformations) automatically.

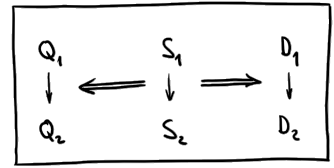

## 5  Evolution of Computational Documents

A different kind of schema evolution is present in programming systems based on documents. The idea, dating back to Boxer [23], is to represent programs as documents with embedded computations. In our model, document can contain simple formulas (akin to those in Potluck [47] or spreadsheets) that may refer to other nodes of the document. We assume that documents can be concurrently edited by multiple users in a local-first manner, i.e., local edits are later synchronized with peers.

The challenge focuses on collaborative conference planning, illustrated in Figure 3 and inspired by a typical use of Notion [49]. A number of co-organizers are planning a conference that will feature talks by several invited speakers. They need to agree on the speaker list, contact speakers and calculate the conference budget. The ideas in this section also apply to computational scientific notebooks that combine code and data [41]. We hope to contribute to ongoing work advancing the format [18, 29, 48].

### Challenge #4: Structured Document Edits

The conference organizers first need to agree on a list of speakers to invite, coordinate who contacts whom and track the acceptance of invitations. They start with a list of three speakers (not shown). The first challenge is to merge document edits done locally by Jonathan and Tomas and change the list and the structure of the document:

1. Jonathan adds an additional speaker to the list and sorts the list of speakers alphabetically by their first name. This changes data, but not the structure (schema) of the document (Figure 3a).
2. Tomas refactors the list into a table. He splits the single textual value into a name and an email (using a comma as the separator) and adds an additional column for tracking which of the organizers should contact the speaker (Figure 3b). This changes the document structure (an implicit schema).

PROGRAMMING CONFERENCE 2023
**Invited speakers**
- Ada Lovelace, ada@rsoc.ac.uk
- Adele Goldberg, adele@xerox.com
- Betty Jean Jennings, betty@rand.com
- Margaret Hamilton, hamilton@mit.com

**(a)** Jonathan adds a new speaker and sorts the list.

PROGRAMMING CONFERENCE 2023
**Invited speakers**

| Name | Email | Who |
|---|---|---|
| Adele Goldberg | adele@xerox.com | TP |
| Margaret Hamilton | hamilton@mit.com | JE |
| Betty Jean Jennings | betty@rand.com | JE |

**(b)** Tomas refactors the list into a table.

PROGRAMMING CONFERENCE 2023

**Invited speakers**
- Adele Goldberg, adele@xerox.com
- Margaret Hamilton, hamilton@mit.com
- Betty Jean Jennings, betty@rand.com

**Conference budget**
Travel cost per speaker:
$1200
Number of speakers:
=COUNT(/ul[id='speakers']/li)
Travel expenses:
=/dl/dd[0] * /dl/dd[1]

**(c)** Tijs adds formulas to compute the total cost.

PROGRAMMING CONFERENCE 2023
**Invited speakers**

| Name | Email | Who |
|---|---|---|
| Ada Lovelace | ada@rsoc.ac.uk | |
| Adele Goldberg | adele@xerox.com | TP |
| Betty Jean Jennings | betty@rand.com | JE |
| Margaret Hamilton | hamilton@mit.com | JE |

**Conference budget**
Travel cost per speaker:
$1200
Number of speakers:
=COUNT(/table[id='speakers']/tbody/tr)
Travel expenses:
=/dl/dd[0] * /dl/dd[1]

**(d)** Geoffrey adopts changes from all three co-organizers.

■ **Figure 3**   Initial version of the document (not shown) contains a list of three speakers (Goldberg, Hamilton, Jennings). The first challenge involves merging a data edit (Figure 3a) with a schema and data edit (Figure 3b); the second involves merging a schema and data edit (Figure 3b) with edit adding code (Figure 3c). Formulas in the final version (Figure 3d) are updated to match the new structure.

The system should be able to merge the changes and produce a table containing the new data in a tabular format (not shown). The result needs to include the additional speaker created by Jonathan, use the order specified by Jonathan, but use the format defined by Tomas. The newly added speaker should be reformatted into the new format; for the "Organizer" column of the newly created table, the system may use a suitable default value or ask the user interactively.

**Remarks: Requirements and Representation**

It should be possible to automatically merge any two sequences of edits (changing both the document structure and data) and the order in which the edits are applied should also not matter. One approach is to use CRDTs [60] which guarantee order-independent convergence over a restricted set of edit operations. This may not always be possible and the system may ask the user for resolution or, possibly, use a default behavior specified by the user.

### Schema Evolution in Interactive Programming Systems

The schema and data evolution happens through a sequence of edits. Table 3 lists edits appearing in our examples (but not a complete list). A system may use a structure editor that provides high-level commands for triggering those edits. Recognizing user intent from plain text editing of source code may make the challenge harder.

> **Merging Data and Code Edits**  In this example, the *schema* is implicit and represents the structure of the document. It involves two independent edits. In one (top right), the data $D_1$ is changed to $D_2'$, but the schema $S_1$ remains unchanged. In the other (bottom left), the schema $S_1$ and data $D_1$ co-evolve into $S_2$ and $D_2$ and the data is further changed to $D_3$. The challenge is to merge the two edits and obtain jointly modified data $D_3'$ of schema $S_2$. In the next section, we will see the same pattern, but also involving code edits. We assume the *convergence* model in this section, but will discuss the more challenging *divergence* model later in the paper.

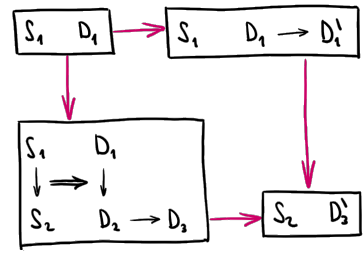

### Challenge #5: Code Co-Evolution for Structured Document Edits

The conference organizers now also use the document to calculate the conference budget. They add a formula that multiplies a fixed value by the number of speakers, obtained by counting the number of element of a specified list in the document.

1. Tijs adds budget calculation to the original document (Figure 3c). This includes two formulas. The first selects all li elements of an ul element with ID `speakers` (not visible in the user interface) and counts their number. The second formula multiplies the constant travel cost per speaker by the number of speakers. Here, we assume the new section uses the HTML definition list `dl` and the formula selects its first and second dd item, respectively.

2. At the same time, Jonathan updates the list of speakers (Figure 3a) and Tomas refactors the data from list into a table as discussed previously (Figure 3b).

The challenge is, again, to merge the edits done independently by the co-organizers. The challenge extends the previous one by adding code. The change to the document structure (*schema*) thus has to co-evolve with both the *data* and the *code*. As shown in Figure 3d, the system needs to change the equation `COUNT(/ul[id='speakers']/li)` to `COUNT(/table[id='speakers']/tbody/tr)`. If the edit is recorded as a list of high-level edits shown in Figure 3 (wrapping element, changing their schema, etc.), then each edit needs to perform corresponding edit to selectors in formulas in the document.

### Remarks: Liveness and Divergence

The above challenge focuses on the local-first software scenario, but it can be extended to also apply to live programming. If formulas in the document are evaluated on-the-fly, the system needs to identify edits that invalidate some of the values (by changing

■ **Table 3** Changes to schema and data and how the programming system should respond

| Change | What | How |
|--------|------|-----|
| Change | Tag of an element | Apply to all matching elements and affected formulas |
| Wrap | Selected element(s) | Wrap elements and update selectors in affected formulas |
| Split | Selected element(s) | Split matching elements using a rule (regular expression) |
| Add | New item to a list | Add data without affecting document structure (schema) |
| Reorder | List items | Reorder data without affecting document structure (schema) |

the data that a formula depends on, or by changing the equation). It should then automatically re-evaluate or invalidate the computed results (requiring explicit re-evaluation). In this scenario, we distinguish between *permanent data* (information in the document) and transient *data* (computed by code). Whereas permanent data should be synchronized and co-evolve with the document structure (schema), transient data can be discarded.

So far, we also assumed that the system follows the *convergence* model, i.e., all variants of the program eventually adopt all changes. A more difficult variant on the challenge is to also support the *divergence* model. In this case, some of the users continue using their variant of the program without adopting all of the changes done by other users. In this model, users may want to adopt all changes done at the *data* layer, but they may selectively choose some of the changes at the *code* and *schema* layers. The challenge is maintaining the same data across multiple program variants and merging changes done to the data based on different document schema. This is the subject of the case studies discussed in the next two sections.

## 6 Divergence Control

Schema evolution becomes more challenging when data, code, and schema cannot all be updated together in an atomic way. In these situations, different versions must coexist and interoperate with one another. For example, edits done to data of a new schema must be applied to data of an older schema, or code written against older schema must run against data of a newer schema.

Such situations are common in software engineering. Web backend architectures often put data and code on separate machines, which makes it impossible to update both atomically. To perform upgrades, teams must manually perform multi-step zero-downtime deployment processes [17] to migrate the database and the code servers forward gradually such that they are compatible with one another at any given step. One example would be deploying a change to add a database column before deploying the code that uses the column.

In the simplest cases, divergence can last only for a brief moment in between zero-downtime deployment steps. But divergence can also occur across longer timescales, especially when the different parts of a system are not all centrally controlled. For example, a public web API may need to support applications using old versions of the API for months or years, since it is not realistic to force all consumers to upgrade immediately. Divergence can even be indefinite. Cambria [46] addresses collaborative applications where different users may want to collaborate on shared data through different tools using different schemas, with no intention of ever converging.

## Challenge #6: Divergence Control for Extract Entity

As a concrete example, we add the problem of divergence control to the earlier Extract Entity challenge. Every month the orders department sends its spreadsheet to the accounting department, which copies the data to its own version of the spreadsheet. This version uses the same structure (schema), but includes additional code for financial tracking. When the orders department migrates its spreadsheet to a database and extracts out customers, the accounting department does not want to conform. They want to continue using their old spreadsheet with their additional code.

The accounting department could manually convert incoming data in the new schema into the old schema. But they shouldn't have to. It should be possible to run new data through the schema change in reverse, converting it back into the format the accounting department is used to ingesting. Note that this is not a matter of synchronizing the divergent spreadsheets – they maintain differently evolved schema.

> **Bidirectional Data Migration**    The scenario adds the need for transferring data changes from a new schema to an older schema. In the diagram on the right, we only show data and schema, ignoring the code added by the accounting department. We start with data $D_1$ of schema $S_1$. 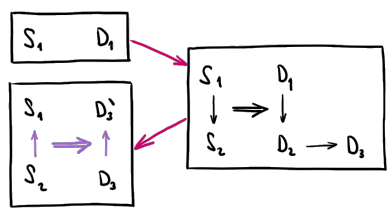
> The accounting department evolves the schema (via Extract Entity), obtaining data $D_2$ of schema $T_2$ using an automatically generated forward transformation. It further modifies the data, obtaining $D_3$. When the accounting department receives data $D_3$, it needs to migrate it back from schema $S_2$ to $S_1$ using an automatically generated backward transformation. That is, schema evolution needs to be accompanied with both forward and backward projection for data (and code) migration.

What is needed is a user-friendly mechanism for transferring data changes from one schema to another bidirectionally. In one direction we want to transfer changes made in the new schema into a variant of the old schema. The opposite direction might be needed, for example, if the accounting department corrects customer addresses, which ought to be pushed through into the orders department's new schema.

## Remarks: Divergence in End-User Workflows

Divergence does not just occur in software engineering; it also appears in end-user workflows. Users frequently make copies of documents like spreadsheets and then diverge them by altering both data content and schema. In the case of spreadsheets schema changes include rearrangements to the structure of rows and columns, while code changes affect formulas. The users then want to transfer such schema, code and data changes between divergent copies, and they want to pick and choose which of the changes to transfer. In practice such transfer is done manually through copy & paste. Basman [7] has documented an ecology of emailed spreadsheets. Burnett et al. [11] distilled field observations of these practices in the story of Frieda:

> *[Frieda is] an office manager in charge of her department's budget tracking. (. . . ) Every year, the company she works for produces an updated budget tracking spreadsheet with the newest reporting requirements embedded in its structure and formulas. But this spreadsheet is not a perfect fit to the kinds of projects and sub-budgets she manages, so every year Frieda needs to change it. She does this by working with four variants of the spreadsheet at once: the one the company sent out last year (we will call it Official-lastYear), the one she derived from that one to fit her department's needs (Dept-lastYear), the one the company sent out this year (Official-thisYear), and the one she is trying to put together for this year (Dept-thisYear).*

> *Using these four variants, Frieda exploratively mixes reverse engineering, reuse, programming, testing, and debugging, mostly by trial-and-error. She begins this process by reminding herself of ways she changed last year's by reverse engineering a few of the differences between Official-lastYear and Dept-lastYear. She then looks at the same portions of Official-thisYear to see if those same changes can easily be made, given her department's current needs. (. . . ) She can reuse some of these same changes this year, but copying them into Dept-thisYear is troublesome (. . . ). Frieda has learned over the years to save some of her spreadsheet variants along the way (. . . ), because she might decide that the way she did some of her changes was a bad idea, and she wants to revert to try a different way she had started before.*

**Maintaining Divergent Variants**  The extra complexity in the above scenario arises from the fact that both schema and code are forked and evolve independently. In one branch, original schema $S_1$ evolve into $S_2$ and original code $C_1$ into $C_3$. In another branch, original schema $S_1$ evolve into $S_2'$ and original code $C_1$ into $C_3'$. The user then wants to merge the independently done changes into a new version of schema and code, written as $S_2 \diamond S_2'$ and $C_3 \diamond C_3'$. The diverging transformations may make conflicting changes that would need to be resolved with assistance from the user.

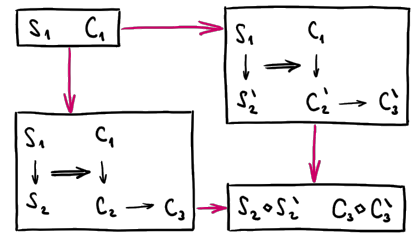

## Remarks: Source Code Version Control

Bidirectional transfer of changes between divergent copies is similar to source code version control as in Git [12]. There is *forking* of long-lived divergent copies. We want to *diff* these forks to see exactly how they have diverged. We want to partially *merge* them by *cherry picking* certain differences. Yet there are also many dissimilarities with Git: our data is more richly structured than lines of text; our schema changes are higher-level transformations on these structures than inserting and deleting characters; and we expect that end-users be able to understand it [22]. The Divergence Control challenge is in a sense to provide "version control for schema change" meeting these criteria, with the key technical challenge being the ability to transfer selected differences bidirectionally through schema changes as independently as possible.

This challenge envisions a new feedback channel that end users have not experienced before: the ability to frictionlessly exchange variations and elaborations they make to documents, replacing manual copy & paste. But meeting this challenge may require a change of perspective in both live programming and local-first software. Live programming must move from being a solitary activity to a collaborative workflow, and further where the collaboration is not just on editing source code but on live integrated code and data, as in the Smalltalk/Lisp images of old. Local-first software must move from automatically converging data replicas to also interactively managing long-term divergence. That implies either application state is being directly exposed to the user, or the application itself offers version control affordances to the user. We encourage both communities to think outside the box of Git, which has proven to baffle not only end-users but also a substantial fraction of developers. The Divergence challenge may be as much a technical challenge as a conceptual challenge, asking us to reconsider the entrenched boundaries between technologies in the stack.

## 7 Multiplicity Change in Data Formats

The challenge of maintaining multiple divergent schemas in programming systems can be well illustrated using *multiplicity change*, which is a common kind of schema change. In this case a field changes from storing a single value to storing multiple values. For example, we might start by storing a single address for each contact in an address book, before realizing that we need to store multiple addresses for a single contact. Alternatively, we might assign each todo in a list to one person before realizing we want the ability to assign a todo to multiple people.

In a relational schema, solving this problem might require extracting a normalized table for the linked entity; some of the challenges of this approach are covered in the Extract Entity challenge. In this section we will instead consider a schema structure with support for arrays; in this context, we can have a multiplicity change by turning a scalar value into an array value. A change of multiplicity in the schema requires producing a transformation for both the corresponding data and the corresponding code. The challenge becomes particularly difficult to handle when there is ongoing divergence between multiple versions of the schema and data.

### Challenge #7: Multiplicity Change in Schema

Consider the following schema for a todo list item, in which each item has a single assignee, represented by a user ID:

```
type Item = { id : id; title : string; assignee : string }
```

Now, we change the schema so that each item has a list of assignees:

```
1  type Item = { id : id; title : string; assignees : Array<string> }
```

Our goal is to preserve the ability for actors in the system to read and write to a shared todo list in either the old or the new schema. For example, an actor should be able to write to either the scalar assignee field when using the original version or to the list of assignees when using the new version.

A natural invariant to preserve across these schemas is that the value of the scalar field should equal the first element of the list field (and if the list is empty, the scalar field should be null). If we were to write code for a one-time data migration from the scalar to list schema, we could easily satisfy this invariant:

```
1  item.assignees = if (item.assignee == null) then [] else [item.assignee]
```

Ongoing edits to the array schema can also be handled in a straightforward way. After edits are made, the value of the scalar field should be set to the first element of the new array (or null if the new array is empty.)

However, handling edits to the scalar schema presents more of a challenge. Consider the following todo with two assignees, presented in terms of the scalar and array schemas. A write is made in the scalar schema to set the new assignee to C.

```
1  todoScalar = { id: 1, title: "Foo", assignee: A }
2  todoArray = { id: 1, title: "Foo", assignees: [A, B] }
3
4  todoScalar.assignee = C
```

What should the array value become after this edit to the scalar schema? To satisfy our invariant, we know that C must become the first element of the array, but this leaves open several options with different tradeoffs:

1. [C]: *Only C should be assigned.* This option produces an array that corresponds directly to the resulting scalar. But it has the downside of deleting data that wasn't even visible in the scalar schema.

2. [C, B]: *Replace A with C.* If the writer wanted to remove A from the assignment and add C, this option performs that intent. However, there is data remaining in the list which was not visible to the scalar writer.

3. [C, A, B]: *Add C to the list.* Perhaps the scalar writer wanted to add C to the assignment without removing anyone; this option satisfies that intent. But it preserves data not visible to the scalar schema.

Because the intent of the writer to the scalar schema cannot be unambiguously interpreted from the write alone, it is impossible to make a perfect choice among these options. A programming system could consult the user to disambiguate their intent, as well as choose (and explain) a default behavior.

> **Bidirectional Data Migration (Revisited)** The diagram illustrating the scenario remains the same as in the case of divergence control for Extract Entity. The initial data and schema $D_1, S_1$ co-evolve into $D_2, S_2$ (adding multiplicity) and a new value is then assigned to the array of assignees, resulting 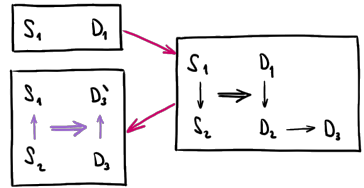
> in $D_3$. The difference from the earlier case is that the backward migration (from $D_3$ to $D_3'$) generated for the schema evolution (from $S_1$ to $S_2$) cannot be generated fully automatically. In case of Extract Entity this was in principle possible.

### Remarks: Managing Design Tradeoffs

If schema, code and data can all be updated atomically together, then multiplicity change is relatively straightforward to handle. Existing scalar data can be trivially migrated to a list, and the code using the data will need to change to accommodate the list data. However, ongoing divergence makes multiplicity change much more difficult, and exposes some general challenges. Different schema may expose partial information, and writers using those schema have to operate without total knowledge of the information available in other schema. As a result, synchronizing writes across the schema requires making difficult tradeoffs, such as choosing to either preserve or destroy hidden data not visible to the writer.

Although there is likely no silver bullet to navigate these tradeoffs, a good solution would give developers or users tools to manage these tradeoffs. For example, a system might allow developers to specify the desired behavior for a particular pair of schemas based on the requirements of the domain. Or a system might even allow users to manually disambiguate their intent for a given write. Whatever the solution, it must be easy to learn and simple to use otherwise it won't be used in practice.

## 8    Live Modeling Languages

We considered live programming in the context of Elm in Section 3, where the challenge was to migrate run-time state (data) to a new schema to keep the program running. Work on language modeling offers a new perspective on the problem. In the context of executable domain-specific modeling languages [40], one can see a program in a domain-specific language as an instance of a metamodel (a static type, or, a schema) that defines the AST structure of the domain-specific language.

A program in a domain-specific language in turn determines (defines/implies/induces) a run-time schema: the run-time structures of code (e.g., classes, inheritance links, declarations, method definitions etc.), as well as the structure of the run-time state (the heap, slots for user data, etc.). Whereas the example discussed in this section is framed in terms of metamodels and models, this can be seen as an expository device: the principles and challenges apply in more grammar-oriented settings. However, the framing also suggests interesting additional challenges.

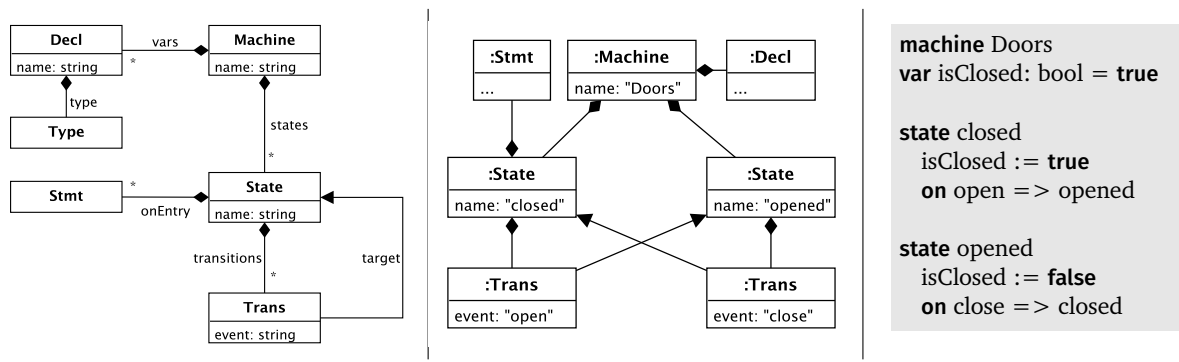

■ **Figure 4**  Definition of the abstract syntax state machine language as a UML metamodel (left) and an example instance, as object diagram (middle) and as code (right).

In the absence of live programming, the run-time *schema* and *code* do not change while running a program in a given domain-specific language and the run-time *state* changes constantly. Live programming, however, requires reconciling changes to the program with the induced run-time structures of both the code and the state. This typically means that at a certain point (a quiescent point) during execution, the code structures need to be updated (hot swapped), and the state needs to be *migrated*.

Figure 4 defines a simple state machine domain-specific language with on-entry actions, and typed global variables (loosely inspired by SML [57]). The figure shows the abstract syntax as UML class diagram (left); an instance modelling the opening and closing of a door in UML object notation (middle) and a possible textual concrete syntax modeling the same state machine (right). The state machine has one global Boolean variable, isClosed, which is flipped when transitioning between the closed and opened states. For the sake of brevity, the abstract syntax model shown on the left omits the details for the on-entry actions (Stmt) and variable types (Type).

In the following we assume that the state machine is executed by an interpreter that processes incoming events, traverses the transitions accordingly, and executes on-entry actions. The notion of the current state and the values of the global variables are the collective run-time state of the program. The structure that the interpreter operates on can be described by a run-time schema. This schema (Figure 5, left) is derived from the abstract syntax schema (Figure 4, left), but is decorated with additional properties and associations modeling the run-time state required for execution. The augmented metamodel includes properties vars (to model variable assignments) and visited (counting number of state visits) to the classes Machine and State, respectively.

An instance of a running state machine consists of an object graph conforming to the run-time schema of a state machine, updating the current state pointer in response to events. This object graph is depicted as a UML object diagram in Figure 5 (right), excluding the statement objects representing on-entry actions. As the diagram shows, the running machine has a current state (closed) and the isClosed field has a value (true) in the vars map. It is instructive to note that the diagram encodes both static and dynamic aspects of the state machine language.

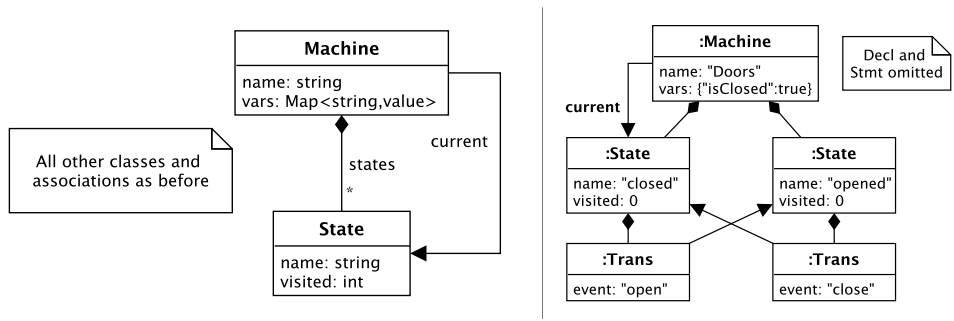

■ **Figure 5**   Augmented run-time metamodel for state machines (left) and a run-time instance of the Doors state machine during execution (right).

## Challenge #8: Live Editing Domain-Specific Languages

The challenge is editing the state machine (Figure 4, right) created in the domain-specific language. This requires to *patch* [64] the run-time structure (Figure 5, right) without discarding and recreating it, possibly requiring migration of run-time state.

The space of possible changes to a state machine is summarized in Table 4. The first column indicates the change category, the second the affected kind of object, and third a description of how to deal with the respective change category. In summary:

- All rows with "simple" in the third column only require a quiescent [70] point in the interpreter loop to update the run-time structure (e.g., as shown in Figure 5).

- Removing a state, however, is only *structurally* simple, since removal is easy, but special care is required if the subject of removal is the current state. In this case, heuristic or user input is needed, such as: reject the edit, point the current state to the initial state, or some other strategy (e.g., the nearest state, previous state etc.)

- Removing a transition or changing the event of a transition potentially has to deal with pending events (e.g., in an event queue) expecting such transitions, since such events are now potentially stale. Strategies to deal with this situation include dropping the events, or preventing the edit when there are pending events.

■ **Table 4**   Possible program changes and how to deal with them at run time

| Change | What | How |
| --- | --- | --- |
| Add/Rename | state | Simple |
| Remove | state | Structurally simple, but heuristic if state is current |
| Add | variable | Add entry to map and initialize |
| Remove | variable | Remove from the variable map (assuming it is unused) |
| Rename | variable | Rename in class, preserving value |
| Type change | variable | Migrate value in map |
| Add | transition | Simple |
| Remove | transition | Structurally simple, but heuristic for pending events |
| Event change | transition | Structurally simple, but heuristic for pending events |
| Add/remove | statement | Hotswap code at quiescent point of interpreter |

- Changing the type of a variable requires a strategy for the current value: either discard and reinitialize, or perform value conversion.
- Note finally that rename variable is mentioned explicitly as a change: this makes it possible to preserve the run-time value of the variable.

**Code as Data Migrations** We could see this challenge as schema, code and data co-evolution as in the case of the Elm architecture in Section 3, but a more interesting perspective is to see the state machine definition as data $D_0$ conforming to a type $S$ defined by the metamodel.

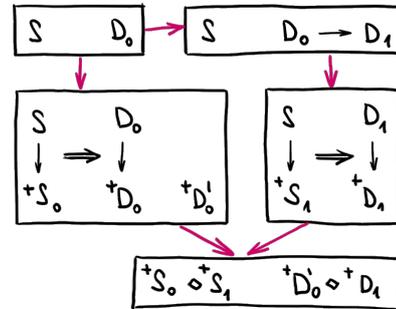

When the state machine is instantiated, the metamodel and the definition are used to derive the run-time type $^+S_0$. The run-time data $^+D_0$ conforming to this type still contain the state machine definition, but also the necessary run-time state required for its execution, which will produce new versions of the run-time data such as $^+D_0'$.

If the state machine definition changes from $D_0$ to $D_1$ (top right), we can synthesize a new run-time state machine type $^+S_1$ and new data $^+D_1$ conforming to it. The challenge is to merge the two run-time types instantiated from different definitions to produce a type $^+S_0 \diamond +S_1$, but also merge the run-time state of the running machine with the new instantiated state to obtain $^+D_0' \diamond {}^+D_1$.

### Remarks: Requirements and Extensions

The goals for this challenge are twofold: from the end-user perspective and from the language engineering perspective. A challenge from the end-user perspective is again to "do minimal harm": it is essential for fluid programming experience that the automatically triggered migrations result in a state as near as possible to the previous application state, to not surprise or confuse the user/developer.

From the language engineering perspective the goal is to employ techniques, formalisms, and tools, to make the construction of such languages easier. The above example is derived from earlier work [57], where the authors manually implemented the run-time patch operation and concluded that even for such a small language (simpler than the example here) it is a complex error-pone endeavor. Furthermore, the field of software language engineering studies and develops generic and reusable techniques to improve the development of DSLs and programming languages, e.g., in the context of language workbenches [25]. The development of live programming languages, however, is currently out of reach for all existing language workbenches.

The above case study is concerned with changes to the DSL program (e.g., the Doors state machine). However, there is another level of schema change to be considered: what if the metamodel (cf. Figure 4) itself is edited? In this case both the programs (e.g., Doors) possibly need to be migrated (in case of a change that is not backwards compatible), *and* the running instance, including the interpreter, since its structure is, too, governed by the abstract syntax of the language.

While the above example is arguably simple, the problem becomes more challenging when the programs themselves define data types, classes, records, structures etc. Since possibly many instances (values, objects) of such data types may exist at run-time, these all have to be migrated in such a way that the programmer experience is minimally disrupted, and that the invariants of said data types is maintained. Another extension, tying in with the data-oriented examples above, involves refactorings of data types in a program. Typically, a refactoring should be behavior preserving, but can it also preserve run-time data? The minimal example is the consistent variable rename in Table 4, which should not have any effect on run-time state.

## 9  Related Work

There has been extensive research on coordinating the evolution of interdependent layers of software, much of it guided by schema and types. But the vast majority of these techniques involve writing custom code in some specialized language, rather than through immediate action and feedback in an interactive programming system. Nevertheless it is important to understand these batch techniques to envision more interactive ones.

**Schema Evolution**    Schema evolution has long been a major problem for SQL databases because the SQL Data Definition Language (DDL) has limited capabilities to redefine existing data. It can rename tables and columns. SQLite [61] uses dynamically typed values allowing them to be implicitly converted if the datatype of the column changes. MYSql [35] can reorder columns without destroying their data. But apart from such special cases, SQL databases have no general purpose support for data migration. As a result in practice schema evolution is mostly done by writing custom Data Manipulation Language (DML) code to migrate the data. Such custom code is greatly complicated if it needs to be done without taking the database offline or must be coordinated across multiple shards. In *Refactoring Databases* Ambler et al. [3] offer a comprehensive taxonomy of schema evolution patterns, including typical strategies for online migration and sample SQL implementations.

Bernstein et al. [9] observed in 2007 "There are hardly any schema evolution tools today. This is rather surprising since there is a huge literature on schema evolution spanning more than two decades." There are now more tools available. Some tools such as Liquibase [45] and PlanetScale[36] could be characterized as version control and continuous integration/deployment for schema. They track schema changes and can calculate diffs in the form of SQL DDL statements to convert one schema version to another, but do not help migrate data. Unfortunately comparing schemas can be ambiguous about the intention of changes. For example has a column been renamed or has it been deleted and a new one created? That distinction makes a big difference to the data in that column. EvolveDB[24] addresses this ambiguity by reverse-engineering the schema into a richer data model and tracking the edits to that model within an IDE. This more precise edit history can be used to infer higher level intentions of a schema change, which then generate SQL scripts to evolve the database. EdgeDB [34] resolves ambiguities by asking questions of the developer, with some answers supplying custom migration code in a proprietary query language.

Some tools provide a Domain Specific Language (DSL) to describe schema evolution. Rails Migrations [54] embeds a DSL in Ruby to manage schema migration but is comparable to the capabilities of SQL DDL, often requiring the addition of custom Ruby or SQL code. Schema evolution is related to the classical view-update problem [4] which was approached from a language perspective by lenses [26]. Generalizing from lenses Curino et al. [20] spawned a stream of research on Database Evolution Languages (DEL) by defining a Schema Modification Operator (SMO) as "a function that receives as input a relational schema and the underlying database, and produces as output a (modified) version of the input schema and a migrated version of the database". SMOs can also rewrite queries to accommodate schema changes. Herrmann et al. [30] defined a relationally complete DEL and then extended it into a Bidirectional Database Evolution Language (BiDEL) [31]. BiDEL appears capable of handling the basic requirements of our *Extract/Absorb Entity* and *Multiplicity Change* challenges, though *Split/Merge Entity* is less clear. It provides schema divergence by supporting multiple schema within one database, but would need some extension to handle data divergence.

Wang et al. [73] use program synthesis to rewrite SQL programs to adapt to a schema change. This process is driven by finding possible correspondences between columns in the old and new schema guided by heuristics on textual similarity and confirmed by the existence of a synthesized equivalent program.

Schema evolution is also a problem for NoSQL [58] databases. While such databases are sometimes called "schemaless" in effect that means the schema is left implicit and tools must try to infer it [62, 63]. Litt et al. [46] uses lenses [26] for bidirectional transformation of JSON. Scherzinger et al. [59] define a set of operators like the relational SMOs discussed above. Chillón et al. [13, 14] offer a more comprehensive set of SMOs that may be capable of handling *Extract/Absorb Entity* and *Multiplicity Change* but not *Split/Merge Entity* nor divergence. None of the above approaches to NoSQL evolution have yet extended into updating code or rewriting queries like their SQL cousins.

*Local-first software* [39] adds the divergence dimension to NoSQL evolution, because the code running on different replicas can evolve at different paces. Litt et al. [46] applies lenses [26] to this problem, specifically discussing the *Multiplicity Change* problem. Schema evolution has also been studied for Object Oriented Databases(OODB) [5, 44]. Smalltalk [27] is itself an OODB, persisting all object instances in an "image file" with some evolution capabilities incorporated in the programming environment [27, pp.252-272]. Gemstone turns the Smalltalk image into a production-quality database and accordingly provides a complex schema evolution API [66].

**Type-Driven Transformations**   Programming language theory undergirds research on type-driven transformations. Bidirectional Transformations [21] have been used for data format conversion and view update. Coupled Transformations [8, 16, 19] express transformations coordinated between types and their instances by encoding them into functions on Haskell GADTs built with strategy combinators. Visser [72] extends the encoding to transform queries written in point-free style. Lämmel [43] recapitulates coupled transformations within logic programming which extends it to transform logic programs. It is not clear how bidirectional transformations can handle generating the unique IDs required in some evolutions, nor the information loss and duplication which can arise in some divergence scenarios.

**Model Driven Engineering**  Model Driven Engineering must also handle evolution. Unlike the textual artifacts involved in other domains, models typically assign unique identifiers enabling more precise differencing and mergeing [1]. Models are often themselves modeled by a metamodel, which lifts the evolution problem to the metalevel: models must be migrated through the evolution of their metamodel, analogously to changing the grammar of a programming language. [32]. Cicchetti et al. [15] study metamodel evolution in the context of divergent changes. Vermolen et al. [71] define a DSL for evolving a data model by generating SQL migrations on the backing database. They consider the evolution of OO-style subclass hierarchies which goes beyond most other work. Their running example includes both our *Extract Entity* and *Multiplicity Change* examples, and would make for a good follow-on problem. However they do not address code/query rewriting.

**Migration by Example**  Programming by example has been used to infer schema migration programs from examples specified by the developer [2, 74]. This work deals with the reality that schema live in the database and so examples must be supplied separately, but we wonder why not incorporate examples into a schema definition language? Textual edits to schema in this language would also adapt the examples, which then could be used to infer an abstract migration, thus avoiding the ambiguities of textually comparing pure schema.

**Data Wrangling**  The evolution of data formats is a concern in Data Science. The Data Diff tool [65] compares two versions of tabular data and finds a minimal matching transformation. AI assistants [52] provide a mechanism for interactively correcting transformations inferred by Data Diff and similar automatic tools. Wrangler [38] observes interactive manipulation of sampled data to write data transformation programs. Petersohn et al. [51] present an algebra of operations on Python dataframes that include shape transformations.

**Live Programming**  Live programming [50, 55, 67] seeks to speed up feedback in (primarily solo) programming. *Hot reloading/swapping* [6, 33] installs code changes into a running system preserving some runtime state but not attempting to migrate that state through type changes. Storm [64] proposes a research agenda for thoroughly live DSL programming focusing on *Semantic Deltas* that unify edit-time and run-time change. Tikhonova et al. [69] study the live modeling problem in Section 8. They use constraints supplied by the developer to solve for the best new runtime state following a DSL language evolution. Rozen et al. [57] take an alternative approach to the same problem using *origin tracking* in textual editing.

## 10    Conclusions

We believe that effective future programming systems will need to tighten the feedback loop, both in making programming more live and in making it more collaborative. One of the key obstacles to making progress in this direction is the problem of *schema evolution*. When the structure of a program evolves, code and data need to co-evolve to match the new structure. The problem of schema evolution has been explored in a wide range of sub-fields, but rarely in its full complexity. Moreover, different sub-fields often use different language, making knowledge transfer between them difficult.

In this paper, we developed a conceptual framework that made it possible to present six case studies from a wide range of areas in a unified way. The areas included live programming, database schema change, model-driven development, spreadsheet and document-oriented programming. Our work highlights the importance of the problem, provides a new uniform way of thinking about it and provides numerous specific challenges that can be used to evaluate the capabilities of new programming systems that aim to tackle the problem.

We did not present new solutions to the problem, but we hope to inspired the reader to do so. Indeed, we look forward to novel solutions to the challenges presented in this paper and the resulting more live and collaborative future of programming!

**Acknowledgements**    We are thank reviewers and attendees of the LIVE 2023 workshop who provided invaluable feedback on earlier version of the paper. We are also grateful to the anonymous reviewers of the journal version of the paper whose feedback greatly improved the quality of the final version. The second author has been supported by Charles University grant PRIMUS/24/SCI/021.

## About the authors

**Jonathan Edwards** is an independent researcher working on drastically simplifying programming. He is known for his Subtext series of programming language experiments and his blog at alarmingdevelopment.org. He has been a researcher at MIT CSAIL and CDG/HARC. He tweets @jonathoda and can be reached at jonathanmedwards@gmail.com.

https://orcid.org/0000-0003-1958-7967

**Tomas Petricek** is an assistant professor at Charles University. He is interested in finding easier and more accessible ways of thinking about programming. To do so, he combines technical work on programming systems and tools with research into history and philosophy of science. His work can be found at tomasp.net and he can be reached at tomas@tomasp.net.

https://orcid.org/0000-0002-7242-2208

**Tijs van der Storm** is a senior researcher in the Software Analysis and Transformation (SWAT) group at Centrum Wiskunde & Informatica (CWI) in Amsterdam, and full professor in Software Engineering at the University of Groningen in Groningen. His research focuses on improving programmer experience through new and better software languages and developing the tools and techniques to engineer them in a modular and interactive fashion. Contact him at storm@cwi.nl.

https://orcid.org/0000-0001-8853-7934

**Geoffrey Litt** is a senior researcher at the independent research lab Ink & Switch. Previously, he completed a PhD at MIT CSAIL in the Software Design Group advised by Daniel Jackson. Before that, he spent five years as an early engineer and designer building the edtech startup Panorama Education (YC S13). Contact him at gklitt@gmail.com.

https://orcid.org/0000-0003-0858-5165